\documentstyle[12pt,fullpage,psfig]{article}
\def\<{\langle}
\def\>{\rangle}
\def\d{\partial}
\def\+{\dagger}
\def\Tr{{\rm Tr}}

\begin{document}
\title{QCD at finite isospin density: from pion
to quark-antiquark condensation}
\author{D.T.~Son$^{1,3}$ and M.A.~Stephanov$^{2,3}$
\\ \small\it 
$^1$ Physics Department, Columbia University, New York, NY 10027
\\\small\it 
$^2$ Department of Physics, University of Illinois, Chicago, 
IL 60607-7059
\\\small\it
$^3$ RIKEN-BNL Research Center, Brookhaven National Laboratory, 
Upton, NY 11973}
\date{August 2000
}
\maketitle

\begin{abstract}
QCD at finite isospin chemical potential $\mu_I$ is studied. This
theory has no fermion sign problem and can be simulated on the lattice
using present-day techniques.  We solve this theory analytically in two
limits: low $\mu_I$ where chiral perturbation theory is
applicable, and asymptotically high $\mu_I$ where perturbative QCD
is at work.  At low isospin density the ground state is a superfluid pion
condensate.  At very high density it is a Fermi liquid with Cooper
pairing. The pairs carry the same quantum numbers as the pions.
Motivated by this observation,
we put forward a conjecture that the transition from hadron to quark
matter is smooth. The conjecture passes several nontrivial tests.  
Our results
imply a nontrivial phase diagram in the space of temperature and
chemical potentials of isospin and baryon number.  At asymtotically
large values of $\mu_I$ and small values of baryon chemical potential
the ground state is in a phase similar to
Fulde-Ferrell-Larkin-Ovchinnikov phase.  It is characterized
by a spatially modulated superfluid order parameter $\<\bar u\gamma_5
d\>$ and may be the asymptotic limit of the inhomogeneous pion
condensation phase advocated by Migdal and others.
\end{abstract}

\section{Introduction}

Good knowledge of QCD in the regime of finite temperature and baryon
density is crucial for understanding a wide range of physical
phenomena.  In cosmology, one faces the problem of understanding
how the Universe has evolved through the QCD phase transition at
temperature $T\sim150$ MeV.  Due to the smallness of the baryon
asymmetry, finite-temperature QCD should be sufficient to deal with
this problem.  However, for the physics
of heavy-ion collisions, one needs to know how QCD behaves when both
temperature and baryon chemical potential are finite.  Lastly,
neutron stars require the knowledge of matter in the ``dense''
regime, i.e. at large baryon densities and very low temperatures.
Much less is known about the last two regimes compared to that of
high-temperature baryon-antibaryon-symmetric QCD.

Let us use neutron stars as an example to illustrate the range of
questions one would like to have answers to.  The equation of state
(EOS) of nuclear matter at high densities determines the
mass-radius relationship and the maximum mass of neutron stars.
Walecka model of nuclear matter predicts that the EOS becomes 
stiffer at higher
densities and approaches the Zel'dovich limit, $\epsilon=p$ (where the
velocity of sound approaches light speed) at very high densities.
However, at asymptotically high densities one expects nuclear matter
to become a weakly-interacting quark liquid, with a much softer
EOS $\epsilon=3p$.  At what density does the transition happen, and
is it a phase transition or a crossover?

Migdal \cite{Migdal} and others \cite{picond} suggested that at very
high densities {\em pion condensation} might happen.  It is also argued
that at even higher densities kaons are condensed \cite{Kcond}.  One
would like to know whether pion and kaon condensations do indeed occur
in nuclear matter, before the transition to quark matter has happened.

Finally, there is a strong recent interest in the phenomenon of {\em
color superconductivity} \cite{colorSC,colorSC2}.  One very
interesting prediction is that at high enough chemical potential, the
ground state of QCD is the ``color-flavor-locking'' state \cite{CFL},
which breaks chiral symmetry.  However, while reliable results can be
obtained at asymptotically high densities, where the strong coupling
is small \cite{Son:1999uk,Rockefeller,inverse}, it is not known how to
extend these results to the region of smaller, more realistic,
densities without having to rely on uncontrollable approximations.

Lacking reliable analytical means to approach QCD in the strong
coupling regime, one naturally turns to numerical methods.  First
principle lattice numerical Monte Carlo calculations provide a solid
basis for our knowledge of the finite temperature regime.  However,
the regime of finite baryon chemical potential $\mu_B$ is still
unaccessible by Monte Carlo, because present methods of evaluating QCD
partition function require taking a path integral with a measure which
includes a {\em complex} fermion determinant.  At zero chemical
potential one can simply ignore the determinant (as one does in the
popular quenched approximation) and still find reasonable results for
physical quantities. However, at finite $\mu_B$ this procedure leads to
qualitatively unacceptable answers as was realized long time ago
\cite{quenched}.  It has been understood more recently
that the quenched approximation breaks down at finite $\mu_B$ because it
describes an unphysical theory containing, beside the normal quarks,
the so-called conjugate quarks with opposite baryon charge
\cite{conjquarks}.

As a side remark, one notes that though the conjugate quarks are
absent in real QCD, there are many theories where they are naturally 
present.
One class of such theories is QCD with two colors, where quarks are
self-conjugate \cite{2color-lattice}.  Another class contains theories
with quarks in the adjoint color representation \cite{2color}.  In all
these theories, the positivity of the fermion determinant ensures the
applicability of lattice Monte Carlo methods.  However, the particle
content of all these theories is very different from the real world.

The failure of the quenched approximation in real QCD at finite
$\mu_B$ and our inability to include complex fermion determinant in a
Monte Carlo simulation is one of the main reasons for our
understanding of QCD at finite baryon density to be still rudimentary.

One way QCD at finite baryon density is different from
finite-temperature QCD is that the transition from hadronic to quark
degrees of freedom occurs due to the large density of a {\em
conserved} charge (such as baryon number) while temperature plays no
role.  This is the motivation for us to turn to QCD at finite chemical
potential $\mu_I$ of {\em isospin\/} (more precisely, of the third
component of isospin, $I_3$), which is conserved by strong interaction.

Before going into details, we would like to comment on the relevance of this
regime to the real world.
Nature does provide us with non-zero $\mu_I$ systems  in
the form of isospin-asymmetric matter (e.g., inside neutron stars), 
however, the latter contains
both isospin density {\em and\/} baryon number density.  In contrast,
the idealized system considered in this paper does not carry baryon
number: the chemical potentials of the two light quarks, $u$ and $d$,
are equal in magnitude, $|\mu_I|/2$, and opposite in sign.  Such a
system, strictly speaking, is unstable with respect to weak decays which do
not conserve isospin, and, as we shall see, is also not electrically neutral
and thus does not exist in the thermodynamic limit.
However, since we are interested in the
dynamics of strong interaction alone, one can imagine that all
relatively unimportant
electromagnetic and weak effects are turned off.  Once this is done, 
we have a nontrivial regime which, as we shall see, is
accessible by present lattice Monte Carlo methods, while being
analytically tractable
in various interesting limits.  As a result, the system we consider has 
a potential to 
improve substantially our understanding of cold dense QCD.  This regime 
carries
many attractive traits of two-color QCD~\cite{2color-lattice,2color}, 
but is realized
in a physically relevant theory --- QCD with three colors.

\section{Positivity and QCD inequalities}

Since in Euclidean space the fermion determinant of our theory is real
and positive, some rigorous results on the low-energy behavior can be
obtained from QCD inequalities \cite{ineqs,2color}.  Let us recall how
the inequalities are derived in vacuum QCD.  The starting point is the
following property of the Euclidean Dirac operator ${\cal D} =
\gamma\cdot (\d + iA) + m$:
\begin{equation}
  \gamma_5 {\cal D} \gamma_5 = {\cal D}^\+ . \label{5d5}
\end{equation}
which, in particular, implies positivity of the determinant, $\det
{\cal D}\ge0$.  For the correlator of a generic meson
$M=\bar\psi\Gamma\psi$, we can write, using (\ref{5d5}) and
Bunyakovsky-Schwartz inequality:
\begin{eqnarray}
  & & \< M(x) M^\+(0) \>_{\psi,A} =
  -\< {\rm Tr} {\cal S}(x,0)\Gamma {\cal S}(0,x) \overline{\Gamma}\>_{A}
  \nonumber\\
  && = \< {\rm Tr} {\cal S}(x,0) \Gamma i\gamma_5 
  {\cal S}^\+(x,0) i\gamma_5\overline{\Gamma} \>_{A} \le 
  \< {\rm Tr} {\cal S}(x,0) {\cal S}^\+(x,0) \>_{A},
\label{corr}
\end{eqnarray}
where ${\cal S}\equiv {\cal D}^{-1}$ and
$\overline{\Gamma}\equiv\gamma_0\Gamma^\+\gamma_0$. The inequality is
saturated for mesons with $\Gamma=i\gamma_5\tau_i$, since ${\cal D}$
commutes with isospin $\tau_i$, which means that the pseudoscalar
correlators majorate all other $I=1$ meson correlators.%
\footnote{It is important, as is the case for $I=1$, that there 
is no disconnected piece after $\psi$ integration in (\ref{corr}). 
The proof does not apply to $\sigma$-meson correlator, $\Gamma=1$.}
As a consequence, one obtains an important restriction on the pattern
of the spontaneous symmetry breaking. For example, the symmetry breaking 
cannot be driven by a
condensate of $\<\bar \psi\gamma_5\psi\>$. Indeed, broken axial SU(2) 
symmetry generators acting on such a pseudoscalar condensate
would have produced $0^+$ Goldstone bosons 
$\bar\psi\tau_i\psi$.

At finite isospin density, $\mu_I\neq0$, positivity still holds
\cite{AKW} and certain inequalities can be derived (in contrast with
the case of $\mu_B\ne0$ when there is no positivity and hence no
inequality can be derived).  Now ${\cal D} = \gamma\cdot (\d + iA) +
{1\over2}\mu_I\gamma_0\tau_3 + m$, and Eq.\ (\ref{5d5}) is not true
anymore, since the operation on the r.h.s.\ of (\ref{5d5}) changes the
relative sign of $\mu_I$.
However, provided $m_u=m_d$, interchanging up and down quarks
compensates for this sign change (the $u$ and $d$ quarks play the role
of mutually conjugate quarks \cite{conjquarks}), 
i.e,
\begin{equation}
  \tau_1 \gamma_5 {\cal D} \gamma_5 \tau_1 = {\cal D}^\dagger.
\label{t5d5t}
\end{equation}
Instead of isospin $\tau_1$ in (\ref{t5d5t}) one can also use $\tau_2$
(but not $\tau_3$.)  Eq.\ (\ref{t5d5t}) replaces the now invalid
Eq.\ (\ref{5d5}) and ensures that $\det D\ge0$.  Repeating the derivation
of the QCD inequalities using (\ref{t5d5t}) we find that the lightest
meson, or the condensate, must be in channels $\bar\psi
i\gamma_5\tau_{1,2}\psi$, i.e., a linear combination of $\pi^-\sim\bar
u\gamma_5 d$ and $\pi^+\sim\bar d\gamma_5 u$ states.  Indeed, as shown
below, in both two analytically tractable regimes of small and large
$\mu_I$ the lightest mode is a massless Goldstone mode
which is a linear combination of $\bar u\gamma_5 d$ and $\bar
d\gamma_5 u$.

\section{Small isospin densities: pion condensate}

When $\mu_I$ is small, 
chiral perturbation theory can be used to treat the problem.  To
have a rough sense of how small $\mu_I$ should be, we require that no
particles other than pions are excited due to the chemical potential.
This gives $\mu_I\simeq m_\rho$ as the upper limit of the
applicability of the chiral perturbation theory.

For zero quark mass and zero $\mu_I$, the pions
are the massless Goldstone bosons of the spontaneously broken
SU(2)$_L\times$SU(2)$_R$ chiral symmetry.  In reality, the quarks have small
masses, which break this symmetry explicitly.
Assuming equal quark masses, the symmetry of the Lagrangian is SU(2)$_{L+R}$.  
The low-energy
dynamics is governed by the familiar chiral Lagrangian, which is written
in terms of the matrix
pion field $\Sigma\in$ SU(2):
\[
  {\cal L} = {1\over4}f_\pi^2{\rm Tr}[\d_\mu\Sigma\d_\mu\Sigma^\+ 
  - 2m_\pi^2 {\rm Re}\Sigma].
\]
This Lagrangian contains only two phenomenological parameters:
the pion decay constant, $f_\pi$, and the pion mass in vacuum,
$m_\pi$.  We will see that interesting physics occurs at $\mu_I>m_\pi$,
and since $m_\pi\ll m_\rho$, there is a nontrivial range of $\mu_I$ 
where the chiral Lagrangian is a reliable and useful treatment.

The isospin chemical potential further breaks SU(2)$_{L+R}$ down to
U(1)$_{L+R}$.  Its effect can be included into the effective
Lagrangian to leading order in $\mu_I$, without introducing additional
phenomenological parameters.  Indeed, $\mu_I$ enters the QCD
Lagrangian in the same way as the the zeroth component of a gauge
potential \cite{2color}.  Thus the finite-$\mu_I$ chiral Lagrangian is
obtained by promoting the global SU(2)$_L\times$SU(2)$_R$ symmetry
to a local gauge symmetry: gauge
invariance completely fixes the way $\mu_I$ enters the chiral Lagrangian 
\cite{2color}:
\begin{equation}
  {\cal L}_{\rm eff}= 
  {f_\pi^2\over4} {\rm Tr} \nabla_\nu \Sigma\nabla_\nu \Sigma^\+
  - {m_\pi^2 f_\pi^2\over2}
  {\rm Re }{\rm Tr} \Sigma.
\label{leff}
\end{equation}
The covariant derivative is defined as
\begin{equation}
  \nabla_0 \Sigma = \partial_0 \Sigma - {\mu_I\over2}
  (\tau_3\Sigma - \Sigma\tau_3),
\qquad \nabla_i \Sigma = \partial_i \Sigma.
\end{equation}
which follows from the transformation property of $\Sigma$ under
rotations by the isospin generator $I_3 = \tau_3/2$.

Using (\ref{leff}) it is straightforward to determine vacuum
alignment of $\Sigma$ as a function of $\mu_I$ and the spectrum of
excitations around the vacuum.  We will be interested in negative
$\mu_I$, which favors neutrons over protons, as in neutron stars.  The
results are very similar to the two-color QCD at finite baryon density
\cite{2color}.  From (\ref{leff}), one finds the potential energy for 
$\Sigma$,
\begin{equation}
  V_{\rm eff}(\Sigma) 
= {f_\pi^2\mu_I^2\over8}\Tr(\tau_3\Sigma\tau_3\Sigma^\+ - 1)
  - {f_\pi^2m_\pi^2\over2}{\rm Re}\Tr\Sigma\ .
  \label{pot}
\end{equation}
The first term in (\ref{pot}) favors directions of $\Sigma$ which
anticommute with $\tau_3$, i.e., $\tau_1$ and $\tau_2$, while the second
term prefers the vacuum direction $\Sigma=1$.  It turns out that the minima
of (\ref{pot}) at all $\mu_I$ are captured by the following Ansatz:
\begin{equation}
  \overline\Sigma = \cos\alpha+i(\tau_1\cos\phi + \tau_2\sin\phi)
    \sin\alpha\ .
  \label{Ansatz}
\end{equation}
Substituting (\ref{Ansatz}) to (\ref{pot}), one sees that the potential
energy depends only on $\alpha$, but not $\phi$:
\begin{equation}
  V_{\rm eff}(\alpha) = {f_\pi^2\mu_I^2\over4}(\cos 2\alpha - 1) -
    f_\pi^2 m_\pi^2 \cos\alpha\ .
  \label{Valpha}
\end{equation}
Minimizing $V_{\rm eff}(\alpha)$ 
with respect to $\alpha$, one sees that the behavior
of the system is different in two distinct regimes:

(i) For $|\mu_I|<m_\pi$, the system is in the same ground state as
at $\mu_I=0$: $\alpha=0$, or $\overline\Sigma=1$.

This result is easy to understand.  The lowest lying pion state costs a
positive energy $m_\pi-|\mu_I|$ to excite, thus at zero
temperature no pion is excited.  The ground state of the Hamiltonian at such
$\mu_I$ coincides with the normal vacuum of QCD.
The isospin density is zero in this case.

(ii) When $|\mu_I|$ exceeds $m_\pi$ 
the minimum of (\ref{Valpha}) occurs at
\begin{equation}
  \cos\alpha={m_\pi^2\over \mu_I^2}.
  \label{alpha}
\end{equation}
In this regime the energy to excite a $\pi^-$ quantum,
$m_\pi-|\mu_I|$, is negative, thus it is energetically favorable to
excite a large number of these quanta.  Since pions are bosons, the
result is a Bose condensate of $\pi^-$.  If the pions did not interact,
the density of the condensate would be infinite. However, the
repulsion between pions stabilizes the system at a finite value of the
isospin density. 
This value can be found by differentiating the ground state energy
with respect to $\mu_I$:
\begin{equation}
  n_I = 
 -{\partial {\cal L}_{\rm eff}\over\partial\mu_I} =
  f_\pi^2 \mu_I \sin^2\alpha
  =f_\pi^2 \mu_I\left(1-{m_\pi^4\over\mu_I^4}\right).
  \label{ni}
\end{equation}
For $|\mu_I|$ just above the condensation threshold, $|\mu_I|-m_\pi\ll
m_\pi$, Eq.\ (\ref{ni}) reproduces the equation of state of the dilute
non-relativistic pion gas \cite{2color},
\[
  n_I = 4f_\pi^2(\mu_I-m_\pi)
\]
At larger $\mu_I$, $|\mu_I|\gg m_\pi$, the isospin density is linear
in $\mu_I$,
\[
  n_I=f_\pi^2\mu_I, \qquad |\mu_I| \gg m_\pi
\]
From Eq.\ (\ref{ni}) one can find the pressure and the energy density as
functions of $\mu_I$.  The interesting quantity is the ratio between
the two,
\begin{equation}
  {p\over\epsilon} = {\mu_I^2-m_\pi^2 \over \mu_I^2+3m_\pi^2}\ .
\end{equation}
%
%
This ratio starts from 0 at threshold and quickly approaches 1 as one
increases $\mu_I$.  Thus, as far as the chiral Lagrangian is still
applicable, the EOS approaches the Zel'dovich limit of maximal
stiffness at high densities, similar to nuclear matter in Walecka
model.

\begin{figure}
  \centerline{\psfig{file=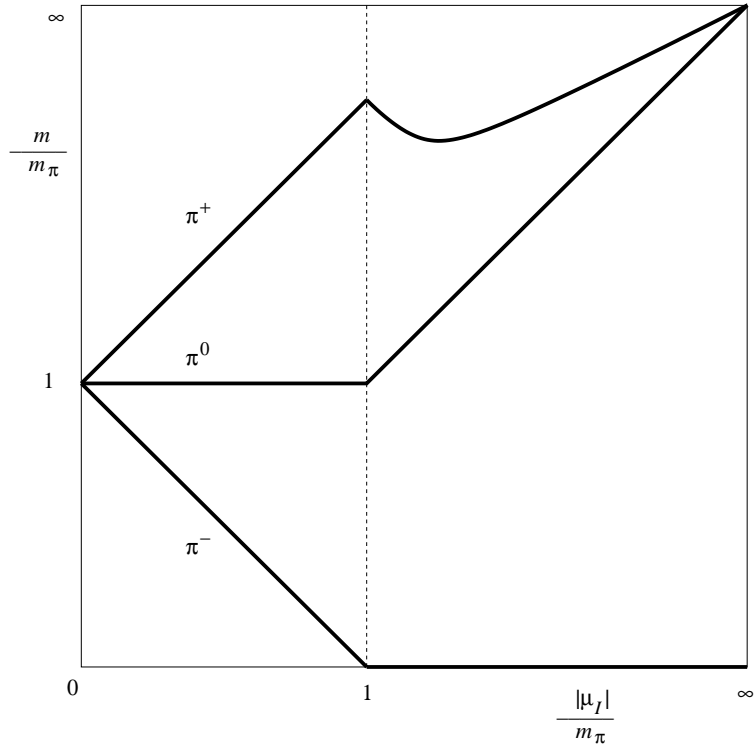,width=.6\textwidth}}
  \caption[]{Schematic plot of masses (rest energies) 
of lowest lying excitations in QCD at finite (negative)
  $\mu_I$, in the regime of applicability of chiral perturbation
theory: $m_\pi,\mu_I\ll m_\rho$.}
\label{fig:excitations}
\end{figure}

The fact that the minimum of the potential (\ref{pot}) is degenerate
with respect to the angle $\phi$ corresponds to the spontaneous
breaking of the U(1)$_{L+R}$ symmetry generated by $I_3$ in the
Lagrangian (\ref{leff}).  This is not unexpected since the ground
state is, in essense, a pion superfluid, with one massless Goldstone
mode.  Since we start from a theory with three pions in vacuum, 
in the superfluid there are, in addition to the massless mode, two
massive modes. One can be identified with $\pi^0$. The other is a linear
combination of $\pi^+$ and $\pi^-$, which we denote as $\tilde\pi^+$,
since it coincides with $\pi^+$ at the condensation threshold.
The mass (defined as the rest energy) of these modes can be obtained
by expanding the Lagrangian (\ref{leff}) around the minimum.  The
result reads (cf. \cite{2color})
\begin{equation}
  m_{\pi^0}= |\mu_I|, \qquad
  m_{\tilde\pi^+}=|\mu_I|\sqrt{1+3(m_\pi/\mu_I)^4}\ .
\end{equation}
At the condensation threshold, $m_{\pi^0}=m_\pi$ and
$m_{\tilde\pi^+}=2m_\pi$, while for $|\mu_I|\gg m_\pi$ both masses approach
$|\mu_I|$ (see Fig.\ref{fig:excitations}).

The values of the chiral  condensate, $\langle\bar u u + \bar d
d\rangle$, and the pion condensate,
$\langle\bar u \gamma_5 d\rangle$ follow from (\ref{alpha}):
\begin{equation}
\langle\bar u u + \bar d d\rangle 
= 2\langle\bar\psi\psi\rangle_{\rm vac} \cos\alpha
\quad\mbox{ and }\quad
\langle\bar u \gamma_5 d\rangle + {\rm h.c.}
= 2\langle\bar\psi\psi\rangle_{\rm vac} \sin\alpha,
\end{equation}
i.e., the chiral condensate ``rotates'' into the pion condensate as 
a function of $|\mu_I|$.

It is also possible to find baryon masses, i.e.\ the energy cost of
introducing a single baryon into the system.  The most interesting
baryons are those with lowest energy and highest isospin, i.e.\
neutron $n$ and $\Delta^-$ isobar.  There are two effects of $\mu_I$
on the baryon masses.  The first comes from the isospin of the
baryons, which effectively reduces the neutron mass by
${1\over2}|\mu_I|$, and the $\Delta^-$ mass by ${3\over2}|\mu_I|$.  If
this was the only effect, the effective $\Delta^-$ mass would vanish
at $|\mu_I|={2\over3} m_\Delta$.  For larger $\mu_I$, baryon or
antibaryon Fermi surfaces would form, which lead to a nonzero baryon
susceptibility $\chi_B\equiv\partial n_B/\partial \mu_B$.  However,
long before that another effect turns on: the $\pi^-$'s in the
condensate tend to repel the baryons, lifting up their masses.

These effects can be treated in the framework of the baryon chiral
perturbation theory \cite{Georgi}. For example, the (Euclidean)
Lagrangian describing nucleons and their interactions with the pions 
at finite $\mu_I$ can be written as:
\begin{equation}
{\cal L}_N =
\bar N \gamma_\mu\nabla_\mu N  
+ m_N \left( \bar N_L\Sigma N_R + {\rm h.c.} \right),
\end{equation}
where
\[
\nabla_0 N = \left(\partial_0 - \frac{\mu_I}2 \tau_3\right) N,
\qquad \nabla_i N = \partial_i N. 
\]
Diagonalizing this bilinear Lagrangian in the pion background
given by $\Sigma=\bar\Sigma$ from (\ref{Ansatz}) one finds
the nucleon masses. The result for the neutron and the $\Delta^-$ 
isobar reads:
\begin{equation}
  m_n = m_N - {|\mu_I|\over2}\cos\alpha,\qquad
  m_{\Delta^-} = m_\Delta - {3|\mu_I|\over2}\cos\alpha,
\label{n,delta}
\end{equation}
in the approximation of nonrelativistic baryons.  
Equation (\ref{n,delta}) can be interpreted as follows: as a result of
the rotation (\ref{Ansatz}) of the chiral condensate, the 
nucleon mass eigenstate becomes a superposition of vacuum $n$ and $p$
states.  The expectation value of the isospin in this state is
proportional to $\cos\alpha$ appearing in (\ref{n,delta}).  With 
$\cos\alpha$ given in Eq.(\ref{alpha}), we see that the two
mentioned effects cancel each other when $m_\pi\ll |\mu_I| \ll
m_\rho$.  Thus the baryon mass never drops to zero, and $\chi_B= 0$
at zero temperature in the region of applicability of the chiral
Lagrangian.

As one forces more pions into the condensate, the pions are packed
closer and their interaction becomes stronger.  When $\mu_I \sim
m_\rho$,
the chiral perturbation theory breaks down.  To find the equation of
state in this regime, full QCD has to be employed.  As we have seen,
this can be done using present lattice techniques since the fermion
sign problem is not present at finite $\mu_I$, similar to the
two-color QCD \cite{2color-lattice}.

\section{Asymptotically high isospin densities: quark-anti\-quark condensate}

In the opposite limit of very large isospin densities, or $|\mu_I|\gg
m_\rho$, the description in terms of quark degrees of freedom applies
since the latter are weakly interacting due to asymptotic freedom.  In
our case of large negative $\mu_I$, or $n_I$, the ground state
contains an equal number of $d$ quarks and $\bar{u}$ antiquarks per
unit volume.  If one neglects the interaction, the quarks fill two
Fermi spheres with equal radii $|\mu_I|/2$.  Turning on the
interaction between the fermions leads to the instability with respect
to the formation and condensation of Cooper pairs, similar to BCS
instability in metals or the diquark pairing at high baryon density
\cite{colorSC}.  To the leading order of perturbation theory, quarks
interact via the one-gluon exchange.  It is easy to see that the
attraction is strongest in the color singlet channel, thus the Cooper
pair consists of a $\bar u$ and a $d$.  The ground state, hence, is a
fermionic superfluid.

The perturbative one-gluon exchange, however, does not discriminate
between the scalar, $\bar u d$, channel, and the pseudoscalar, $\bar u
\gamma_5 d$, channel: the attraction is the same in both cases.  But
one can expect that the instanton-induced interaction, however small,
will favor the $\bar u \gamma_5 d$ channel over the $\bar u d$ one.
The condensate thus is pseudoscalar and breaks parity, 
\begin{equation}
  \langle \bar{u} \gamma^5 d \rangle \ne 0.
  \label{ubard_cond}
\end{equation}
This is consistent with our earlier observation that QCD inequalities
constrain the $I=1$ condensate to be a pseudoscalar at any $\mu_I$.
Note that the order parameter in (\ref{ubard_cond}) has {\em the same
quantum numbers} as the pion condensate at lower densities.  We shall
discuss this coincidence later.

As a consequence of the Cooper pairing, the fermion spectrum acquires
a gap $\Delta$ at the Fermi surface, where
\begin{equation}
  \Delta = 
  b |\mu_I| g^{-5} e^{-c/g}, \qquad c = 3\pi^2/2,
  \label{gap}
\end{equation}
where $g$ should be evaluated at the scale $|\mu_I|$.  The peculiar
$e^{-c/g}$ behavior comes from the long-range magnetic interaction, as
in the superconducting gap at large $\mu_B$ \cite{Son:1999uk}.  The
constant $c$ is smaller by a factor of $\sqrt{2}$ compared to the
latter case due to the stronger one-gluon attraction in the singlet
$q\bar{q}$ channel compared to the $\bar{\bf 3}$ diquark channel.
Consequently, the gap (\ref{gap}) is exponentially larger than the
diquark gap at comparable baryon chemical potentials.  Using the
methods of \cite{Rockefeller} one can estimate $b\approx 10^4$.  As in
the BCS theory, the critical temperature, at which the superfluid
state is destroyed, is of order $\Delta$.

Asymptotically, $\Delta$ is much less than $\mu_I$, and superfluidity
has little effect on the equation of state.  The ratio $p/\epsilon$
approaches $1/3$ from below in the limit $\mu_I\to\infty$.

\section{Quark-hadron continuity and confinement}

Since the order parameter (\ref{ubard_cond}) has the same quantum
numbers and breaks the same symmetry as the pion condensate in the
low-density regime, it is plausible that there is no phase transition
along the $\mu_I$ axis.  In this case, as one increases the density,
the Bose condensate of weakly interacting pions smoothly transforms
into the superfluid state of $\bar{u}d$ Cooper pairs.  The situation
is very similar to that of strongly-coupled superconductors with a
``pseudogap'' \cite{pseudogap}, and possibly of high-temperature
superconductors \cite{Randeria}.  This also parallels the continuity
between nuclear and quark matter in three-flavor QCD as conjectured by
Sch\"afer and Wilczek \cite{continuity}.  We hence conjecture that in
two-flavor QCD one can move continuously from the hadron phase to the
quark phase without encountering a phase transition.  We stress here
that this conjecture needs to be verified by lattice calculations.

At first sight, this conjecture seems to contradict a common
wisdom that there is a ``deconfinement'' phase transition from the
hadron phase to the quark phase.  It is logically possible that there
exists a first order phase transition at intermediate value of $\mu_I$.
However, there are several nontrivial arguments
that make the continuity hypothesis highly plausible.

The first argument arises from considering baryons.  One notices that
all fermions have a gap at large $|\mu_I|$, which means that all
excitations carrying baryon number are massive.  In particular, at
zero temperature, the baryon number susceptibility $\chi_B$ vanishes.
This is also true at small $\mu_I$.  It is thus natural to expect that
all excitations with nonzero baryon number are massive at any value of
$\mu_I$, and $\chi_B$ remains zero at $T=0$ for all $\mu_I$.  This
also suggests one way to verify the continuity on the lattice.

Another argument comes from considering the limit of large number of
colors $N_c$.  Let us recall that in finite-temperature QCD, there is
a mismatch, at large $N_c$, between the number of gluon degrees of
freedom, which is ${\cal O}(N_c^2)$, and of hadrons, which is ${\cal
O}(N_c^0)$.  This fact is a strong hint of a first order
confinement-deconfinement phase transition, at which the effective
number of degrees of freedom jumps from ${\cal O}(N_c^0)$ to ${\cal
O}(N_c^2)$.  It is easy to see, however, that the $N_c$ behavior of
thermodynamic quantities is the same in the ``hadronic'' phase (low
$\mu_I$) and the ``quark'' phase (large $\mu_I$).  Indeed, at very
large $\mu_I$ the isospin density $n_I$ is proportional to the
number of quarks, which is ${\cal O}(N_c)$:
\begin{equation}\label{ni2}
  n_I={N_c\over3}{\mu_I^3\over8\pi^2}.
\end{equation}
In the small $\mu_I$ region the isospin density is given by Eq.\
(\ref{ni}).  At large $N_c$ limit the pion decay constant scales as
$f_\pi^2={\cal O}(N_c)$, and thus the isospin density in the pion gas
is also proportional to $N_c$.\footnote{
With physical values of $N_c$, $f_\pi$ and $m_\pi$, the values of
$n_I$ given by eqs. (\ref{ni}) and (\ref{ni2}), naively
continued into the regime of intermediate $\mu_I$, cross at
$\mu_I\approx 800$MeV. This agrees with the value of
$\mu_I\sim m_\rho$ where one would expect the crossover
between the quark and hadron regimes to occur. This is a
quantitative indication that a phase transition is not
necessary.}  
What happens is that the
repulsion between pions becomes weaker as one goes to larger $N_c$,
thus more pions can be stacked at a given chemical potential.  As a
result, the $N_c$ dependence of thermodynamic quantities is the same
in the quark and the hadronic regimes, although for seemingly very
different reasons.

Now let us return to the question of confinement.  Naively, one would
think that at asymptotically large $\mu_I$, the $\bar{u}$ and $d$
quarks are packed at a very high density, and the system should 
become deconfined.  At finite temperature, there is no rigorous way to
distinguish the confined and deconfined phases in QCD with quarks in
the fundamental representation.  However, at zero temperature (and
finite $\mu_I$), a sharp distinction can be made between the two
phases.  In the confined phase, all particle excitations carry integer
baryon number; the deconfined phase can be defined as the phase where
there exist finite-energy excitations carrying fractional baryon
charge.  The pion superfluid at small $\mu_I$ clearly is in the
confined phase.  The question is: is the quark matter at large $\mu_I$
confined or deconfined?

It might seem that at very large $\mu_I$ there exist excitations
with fractional baryon number.  These are the fermionic quasiparticles
near the Fermi surface, which are related to the original quarks and
antiquarks by a Bogoliubov-Valatin transformation.  The opening of a
BCS gap makes the energy of these excitations larger than $\Delta$,
but still finite.

To see that the logic above has a fault and there are no such
excitations, one needs to consider dynamics of very soft gluons.  The
crucial observation is that at large $\mu_I$, gluons softer than
$\Delta$ are not screened neither by Meissner nor by Debye effect.%
\footnote{
	This is similar to the behavior of the unbroken
	SU(2)$_c$ sector of two-flavor color superconductors
	\cite{Rischke}.
}  
Meissner effect is absent because the condensate does
not break gauge symmetry (in contrast to the color superconducting
condensate \cite{colorSC}).  Debye screening is also absent, because
on scales softer than $\Delta$ there are no charge excitations in the
medium: the Cooper pairs are neutral, while the fermions are too heavy
to be excited.  Thus, the gluon sector below the $\Delta$ scale is
described by pure SU(3) gluodynamics, {\em which is a confining
theory}.  This means there are no quark excitations above the ground
state: all particles and holes must be confined in colorless objects,
mesons and baryons, just like in vacuum QCD.

If there is no transition along the $\mu_I$ axis, we expect
confinement at all values of $\mu_I$.  At large $\mu_I$, 
since the running strong
coupling $\alpha_s$ at the scale of $\Delta$ is small, the confinement
scale $\Lambda_{\rm QCD}'$ 
is much less than $\Delta$.  In more detail, let us imagine following
the running of the strong coupling from the UV to the IR. First,
$\alpha_s$ increases until the scale $g\mu_I$ is reached when it
``freezes'' due to Debye screening and Landau damping.  The freezing
continues until we reach the scale $\Delta$, after which the coupling
runs again as in pure gluodynamics.  Since the coupling is still small
at the scale $\Delta$, it can become large only at some scale
$\Lambda_{\rm QCD}'$ much lower than $\Delta$.  Thus, at large
$|\mu_I|$ there are three different scales separated by large
exponential factors, $\mu_I\gg\Delta\gg\Lambda_{\rm QCD}'$.

That the scale of confinement is much smaller than the gap at large
$\mu_I$ has an important consequence for finite temperature.  One can
actually predict a temperature driven deconfinement phase transition
at a temperature $T_c'$ of order $\Lambda_{\rm QCD}'$.  Indeed, at
such low temperatures, quarks are unimportant, so the transition must
be of first order as in pure gluodynamics.  In particular, one expects
the baryon number susceptibility temperature dependence to change from
$e^{-3\Delta/T}$ to $e^{-\Delta/T}$ around $T_c'$ due to
deconfinement.




The smallness of the confinement scale $\Lambda_{\rm QCD}'$ compared
to the BCS gap $\Delta$ allows one to conclude that the binding energy
of quarks and antiquarks is small and the hadronic spectrum follows
the pattern of the constituent quark model, with $\Delta$ playing the
role of the constituent quark mass.  This means mesons weigh
$2\Delta$ and baryons weigh $3\Delta$, approximately.  A good analog
of the large $\mu_I$ regime is vacuum QCD with only heavy quarks.  As in the
latter case, the string tension and string breaking are determined by
parametrically different energy scales ($\Lambda_{\rm QCD}'$ and
$\Delta$, respectively).  Hence the area law should work up to some
distance much larger $\Lambda_{\rm QCD}^{-1}$, even when fundamental
quarks are present.  For the same reason one also expects the
high-spin excited states of hadrons to be narrow at large $\mu_I$.

The energy hierarchy also leads to a curious dispersions relation of
hadrons in the isospin dense regime.  Consider, for example, the
$\rho^-$ meson, which is a bound state of a $\bar u$ and a $d$.  At zero
total momentum, the $\bar u$ and the $d$ are on the opposite sides of the
Fermi surface.  As one increases the total momentum, the two
constituents move along the Fermi surface, {\em remaining close to the
latter} until the total momentum becomes larger than $\mu_I$,
i.e. twice the Fermi momentum.  Thus, the dispersion curve of the
$\rho^-$ must remain essentially {\em flat} in the interval of
momentum $(0,|\mu_I|)$.  For baryons, energy is almost independent of
the momentum in the interval $(0,1.5|\mu_I|)$.  
The group velocity of
hadrons, thus, almost vanishes in these intervals.  Above these
intervals it should be equal to the speed of light.  
It would be interesting to follow, on the lattice, the evolution of
the dispersion curves of $\rho^-$ from small to large $\mu_I$.

\section{The phase diagrams on $(T, \mu_I)$ and $(\mu, \mu_I)$ planes}

By considering nonzero $\mu_I$, we make the phase diagram of QCD
three-dimensional: ($T$, $\mu_B$, $\mu_I$).  Two planes in this
three-dimensional space are of a special interest: the $\mu_B=0$
$(T,\mu_I)$ plane, which is completely accessible by present lattice
techniques, and the $T=0$ $(\mu_I,\mu_B)$ plane, where the neutron
star matter belongs.

\begin{figure}
  \centerline{\psfig{file=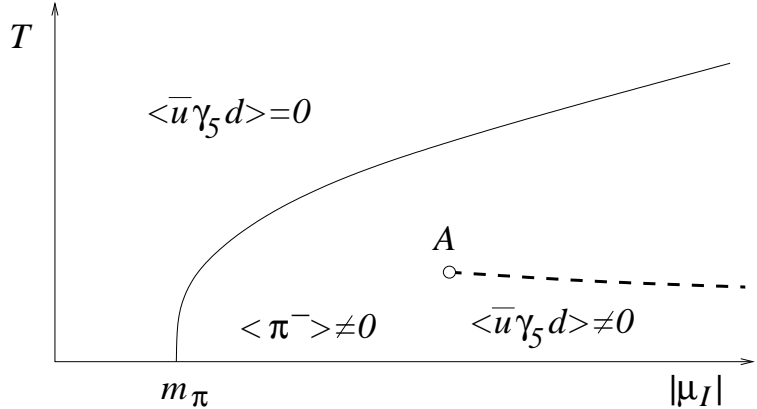,width=.6\textwidth}}
  \caption[]{Phase diagram of QCD at finite isospin density.}
\label{fig:tmui}
\end{figure}

Let us first consider the phase diagram on the $(T, \mu_I)$ plane,
which is simpler.  Two phenomena determine the phase plane on this
plane (Fig.\ref{fig:tmui}): pion condensation and confinement.
At sufficiently high temperature the condensate (\ref{ubard_cond})
melts (solid line in Fig.~\ref{fig:tmui}). For large $\mu_I$, this
critical temperature is proportional to the BCS gap (\ref{gap}). There
are two phases which differ by symmetry: the high temperature phase
where the explicit flavor $U(1)_{L+R}$ symmetry is restored, and the
low-temperature phase where this symmetry is spontaneously broken.
The phase transition is in the O(2) universality
class.\footnote{
	The width of the Ginzburg region is suppressed
	by $(\Delta/\mu_I)^4$ at large $\mu_I$ as in usual BCS
	superconductors
	and also by $1/N_c^2$ at large $N_c$ as at the QCD chiral
					 transition
	\cite{noteGinzburg}.
}  
The critical temperature $T_c$ vanishes at
$\mu_I=m_\pi$ and is an increasing function of $\mu_I$ in both regimes
we studied: $|\mu_I|\ll m_\rho$ and $|\mu_I|\gg\Lambda_{\rm QCD}$.
Thus, it is likely that $T_c(\mu_I)$ is a monotonous function of
$\mu_I$.  In addition, as explained before, at large $\mu_I$, there is
a first-order deconfinement phase transition at some temperature
$T_c'$ much lower than $T_c(\mu_I)$.  Since there is no phase
transition at $\mu_I=0$ (for small $m_{u,d}$) or at $T=0$ (assuming
quark-hadron continuity), this first-order line must end at some point
$A$ on the $(T, \mu_I)$ plane (Fig.~\ref{fig:tmui}).

The phase diagram in the $(\mu_I,\mu)$ plane at zero temperature turns out
to be quite complicated.  We defer the more detail study of this plane
for future work.  Here we shall only consider the regime
$|\mu_I|\gg\mu_B$, both much larger than $\Lambda_{\rm QCD}$, so that
perturbative QCD can be used.  When $\mu_B=0$ and
$|\mu_I|\gg\Lambda_{\rm QCD}$ we have seen that the system is a
superfluid with a gap $\Delta$.  Finite $\mu_B$ provides a mismatch
between $\bar{u}$ and $d$ Fermi spheres. The superconducting state
becomes unfavorable at some value of $\mu_B$ of order $\Delta$.  It is
known \cite{FFLO} that the destruction of this state occurs through
{\em two} separate phase transitions.
As one increases $\mu_B$, at $\mu_B$ slightly below $\Delta/\sqrt{2}$,
a first-order phase transition takes the system to the
Fulde-Ferrell-Larkin-Ovchinnikov (FFLO) state \cite{FFLO}, which is
characterized by a spatially modulated superfluid order parameter
$\<\bar u\gamma_5 d\>$ with a wavenumber of order $2\mu_B$.  How
exactly this spatial dependence looks like is still unknown, mostly because
the FFLO state has not been observed in metals.  The FFLO state
persists only until $\mu_B=0.754\Delta$ when it goes through a
second-order phase transition to a $\<\bar u\gamma_5 d\>=0$ state.
The latter must be a color superconductor with one-flavor diquark
condensates $\<uu\>$ and $\<dd\>$, due to the attraction between
quarks of same flavor.
In the region of the $(\mu_B\mu_I)$ 
phase diagram directly relevant to neutron stars, $\mu_B>\mu_I$, 
the color superconducting FFLO phase is studied in a recent 
paper of Alford, Bower and Rajagopal \cite{loff}.

The most interesting feature of the FFLO state is that it has the same
symmetries as the inhomogeneous pion condensation state which might
form in electrically neutral nuclear matter at high densities, as
argued by Migdal \cite{Migdal} and others \cite{picond}.  The FFLO
phase, thus, can be thought of as a realization of Migdal's pion
condensate in the regime of asymptotically high densities.  It is also
conceivable that the two phases are actually one, i.e., continuously
connected on the $(\mu_I,\mu_B)$ phase diagram.


\section{Conclusion}

Our original and primary motivation for considering QCD at finite
isospin densities is to have a dense regime of realistic, three-color
QCD that can be studied on lattice.  Based on analytical calculations
in the two asymptotic regimes of low and high densities, we found that
there is likely no phase transition along the $\mu_I$ axis at zero
temperature.  This conjecture should be verified on the lattice.  An
obvious way is to study the thermodynamics of the system.  If our
continuity conjecture is correct, all thermodynamic quantities should
be smooth functions of $\mu_I$.  In this case, we also suggest that
the ratio $p/\epsilon$ is a non-monotonic function of $\mu_I$: it
raises from 0 at the threshold $\mu_I=m_\pi$ to some value close to 1,
then drops to some minimal value, and then approaches $1/3$ from
below at large $\mu_I$.  The baryon susceptibility should vanish at
any $\mu_I$ at zero temperature.  We also predict the existence of a
line of first-order phase transition on the $(T,\mu_I)$ plane, which
terminates at a second order point.

The phase diagram on the $(\mu_I,\mu_B)$ plane, which is most relevant
for neutron star physics, remains inaccessible to the lattice.  Based
on our preliminary investigations, the phase diagram on this plane
should have a rather complicated topology.  The most interesting
feature of this diagram appears to be the existence of the FFLO phase,
which is reliably predicted at $\mu_I\gg\mu_B$, both being large.
This phase has the same symmetry as the pion condensation state
conjectured by Migdal, and both might be different regions of a single
connected region on the phase diagram.

The authors thank L.~McLerran, J. Kogut, R.~Pisarski,  E.~Shuryak for
discussions,  K.~Rajagopal for drawing their attention to
Ref.\ \cite{FFLO}, and the DOE Institute for Nuclear Theory
 at the University of Washington for
its hospitality.  The work of DTS is supported, in part, by DOE
grant DE/FG02-92ER40699.

\subsection*{Note Added}

Further research \cite{gcm} reveals that the physics below the
scale $\Delta$ is described by ``gluodynamics of continuous media''
with a large dielectric constant. As a result, the scale of the confinement
$\Lambda_{\rm QCD}'$ is small and decreases exponentially with the chemical
potential $\mu_I$. This means that the line of the first order 
deconfinement transition goes down as indicated in Fig.~\ref{fig:tmui}.

\end{document}